\documentclass[fleqn,10pt]{wlscirep}
\usepackage[utf8]{inputenc}
\usepackage[T1]{fontenc}

\usepackage{longtable}
\usepackage{caption}
\usepackage{graphicx}
\usepackage{float} 
\usepackage{subfigure}
\usepackage{tabularx}
\usepackage{fancyhdr}
\usepackage{stfloats}
\usepackage{tabularray}
\usepackage{booktabs}
\usepackage{ulem}
\usepackage{color,xcolor}
\usepackage{multirow}
\usepackage{graphicx}
\definecolor{mygreen}{RGB}{61,145,64}
\usepackage{threeparttable}

\title{BookGPT: A General Framework for Book Recommendation Based on a Large Language Model}

\author[1]{Aakas Zhiyuli}
\author[2,*]{Yanfang Chen}
\author[3]{Xuan Zhang}
\author[3]{Xun Liang}
\affil[1]{AI for Science Institute, Beijing, China}
\affil[2]{Renmin University of China, Libraries, China}
\affil[3]{Renmin University of China, School of Information, China}
\affil[*]{Corresponding author: cyf@ruc.edu.cn}

\begin{abstract}
With the continuous development and change exhibited by large language model (LLM) technology, represented by generative pretrained transformers (GPTs), many classic scenarios in various fields have re-emerged with new opportunities. This paper takes ChatGPT as the modeling object, incorporates LLM technology into the typical book resource understanding and recommendation scenario for the first time, and puts it into practice. By building a ChatGPT-like book recommendation system (BookGPT) framework based on ChatGPT, this paper attempts to apply ChatGPT to recommendation modeling for three typical tasks, book rating recommendation, user rating recommendation, and book summary recommendation, and explores the feasibility of LLM technology in book recommendation scenarios. At the same time, based on different evaluation schemes for book recommendation tasks and the existing classic recommendation models, this paper discusses the advantages and disadvantages of the BookGPT in book recommendation scenarios and analyzes the opportunities and improvement directions for subsequent LLMs in these scenarios. The experimental research shows the following. (1) The BookGPT can achieve good recommendation results in existing classic book recommendation tasks. Especially in cases containing less information about the target object to be recommended, such as zero-shot or one-shot learning tasks, the performance of the BookGPT is close to or even better than that of the current classic book recommendation algorithms, and this method has great potential for improvement. (2) In text generation tasks such as book summary recommendation, the recommendation effect of the BookGPT model is better than that of the manual editing process of Douban Reading, and it can even perform personalized interpretable content recommendation based on readers' attribute and identity information, making it is more persuasive than interpretable one-size-fits-all recommendation models. Finally, we have open-sourced the relevant datasets and experimental codes, hoping that the exploratory program proposed in this paper can inspire the development of more LLMs to expand their applications and theoretical research prospects in the field of book recommendation and general recommendation tasks.
\end{abstract}
\begin{document}

\flushbottom
\maketitle
%
%
\thispagestyle{empty}

\section*{Introduction}

Book understanding and personalized recommendation (BUPR) are crucial applications in the field of library and information science (LIS). In the BUPR scenario, we typically need to solve several subproblems, such as (1) how to recommend suitable books based on users' interests and preferences, (2) how to predict the popularity of a new book for deciding whether to purchase it, and (3) how to provide interpretable recommendations for different users to improve the user adoption rate. In general, we need to model the interactions between readers and books, readers' basic attributes, and books' basic attributes, among other types of feature and attribute data, and use different machine learning methods to train and optimize independent recommendation models for each subtask, thus improving the final recommendation effect. However, as the scenario becomes increasingly complex and the amount of data grows, it becomes challenging to meet various application and recommendation needs. Is it possible to have a unified personalized recommendation paradigm that enables us to solve all kinds of basic problems in the BUPR scenario with just a few task-relevant training examples? The answer is yes!

In recent years, with the rapid development of natural language processing (NLP) technology, significant changes have emerged, both in terms of the scale of model parameters and the richness of training data. For example, in early December 2022, OpenAI released a chatbot based on GPT-3.5\cite{zhao2023survey} called Chat Generative Pretrained Transformer (ChatGPT) \cite{openaiChatgpt}. This chatbot is based on large-scale pretrained language models and fine-tuned for efficient natural language understanding, as well as logical reasoning in multiturn conversations. Specifically, it can perform a variety of NLP tasks, including assisting with code writing, summarizing documents, and continuing novel writing. Since its release, this model has sparked heated discussion in industry.

\begin{figure}[H]
	\centering
	\includegraphics[width=\linewidth]{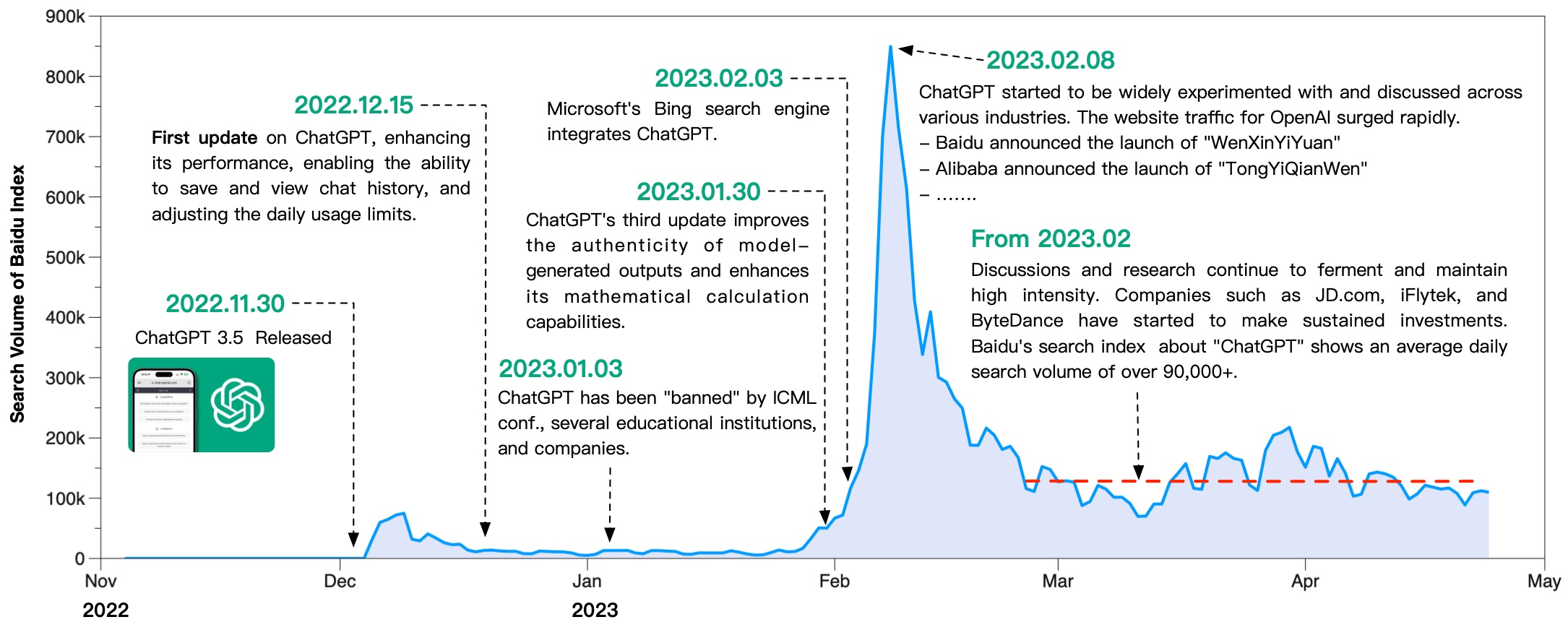}
	\caption{\label{chatgpt_baidu_index} ChatGPT's search volume in the Baidu Index from November 2022 to April 2023.}
\end{figure}

Figure 
\ref{chatgpt_baidu_index} shows the trend chart of the Baidu search index for ChatGPT from its initial release in early December 2022 to the present day, with typical important events annotated. From the chart, we can see that during the initial release stage (November 2022 - February 2023), due to the imperfect model performance and user interface, the overall popularity of ChatGPT remained at a relatively low level. Starting in February, with OpenAI's iteration of the model, as well as the reporting of several important events, such as ChatGPT passing the Google engineer interview \cite{passGoogle} and the extensive participation of many companies, ChatGPT began a thriving journey of research and application. By the end of April, the overall daily search volume for ChatGPT on Baidu reached 90,000. Furthermore, the technical foundations of ChatGPT, its large language models (LLMs), have also received great attention from the academic community \cite{zhao2023survey}.

In particular, in the LIS field, an increasing amount of research has been conducted on the theoretical aspects related to ChatGPT-like models. For example, studies have examined the technical ethics and risks associated with ChatGPT-like models\cite{lund2023chatting,cox2023chatgpt,verma2023novel,zhixiong2023influence}, as well as how to better utilize ChatGPT in LIS\cite{panda2023exploring,kirtania2023openai}. However, these studies mainly focused on theoretical influences and application analyses of ChatGPT without conducting extensive experiments and testing in practical LIS scenarios. It is worth exploring and researching a through experimental verification of whether ChatGPT models can be used to build recommendation frameworks to solve recommendation problems in LIS and how ChatGPT models perform compared to traditional recommendation models in these scenarios, as along with their advantages and disadvantages. Therefore, this paper focuses on the BUPR scenario to empirically investigate the feasibility of using ChatGPT models in book recommendation settings and evaluate the performance of ChatGPT models through experimental comparisons on BUPR tasks.

The main contributions of this article include the following. (1) For the first time, this article applies LLMs such as ChatGPT to build a unified recommendation system framework for the classic BUPR task in the LIS field to explore the possibility of applying LLMs in LIS. (2) Based on the BUPR scenario, we discuss construction ideas and prompt engineering methods for three subtasks: the book rating task, user book rating preference recommendation task, and book summary recommendation task. Two different prompt modelling methods, zero-shot modelling and few-shot modelling, are verified in terms of their feasibility through empirical research. (3) Finally, this article open-sourced the data and testing schemes involved in the experimental process to facilitate further research and discussion on the corresponding issues.

The organization of this article is as follows. First, we discuss the related work involved in this study from two directions: "LLMs" and "book recommendations". Then, we first provide an overview of the BookGPT model, followed by detailed formal definitions of the three subtasks of book recommendation. Then, we further propose the construction methods of prompt engineering, the verification methods applied to the output results, and the methods for evaluating recommendation effectiveness. Afterwards, we conduct a detailed experimental evaluation of the ChatGPT-like book recommendation system (BookGPT), including a dataset analysis, an evaluation scheme design, and an experimental results discussion. Finally, we summarize the research content and focus of this article and provide future research directions in the field of book recommendation based on LLMs.

\section*{Related Works} 

\subsection*{Large Language Models}

LLMs typically refer to NLP models with parameter sizes exceeding billions. Recently, research on LLMs has become an important frontier in the field of NLP, from the widely used and researched statistical language models (SLMs) \cite{jelinek1998statistical,rosenfeld2000two,liu2004statistical}, to neural language models (NLMs) based on neural networks for NLP\cite{bengio2000neural,mikolov2010recurrent}, to pretrained language models (PLMs) \cite{mikolov2013efficient,devlin2018bert,sarzynska2021detecting}, and finally to LLMs \cite{brown2020language,chowdhery2022palm,touvron2023llama,zeng2023glm130b}. With the iteration of these models, NLP technology has exhibited typical characteristics: the model parameter scale is becoming larger, the context awareness of these models is strengthening, the durations of multiturn conversations are lengthening, and multiple modalities of interaction are being utilized.

An LLM is a neural network architecture model based on the transformer mechanism \cite{vaswani2017attention} that extracts and expresses natural language features by introducing multihead attention and stacking multiple layers of deep neural networks. Among the various types of available LLMs \cite{brown2020language, chowdhery2022palm, touvron2023llama, zeng2023glm130b}, the main differences lie in the sizes of their training corpora, model parameter sizes, and scaling sizes. Based on a well-designed prompt engineering strategy \cite{PromptGuide2022}, LLMs trained on large-scale corpora can usually produce good dialogue results. The natural language understanding and response abilities exhibited by current LLM models are generally believed to be emergent abilities \cite{wei2022emergent, jiang2023latent} resulting from the tremendous growth in the number of model parameters and the size of the training corpus; i.e., when the parameter scale exceeds a certain level, the developed model exhibits new abilities that are radically different from those of its previous levels, such as in-context learning (ICL) \cite{lampinen2022can} and chain of thought (CoT) \cite{wei2022chain}.

Popular versions of LLM models currently include the GPT3/4 series of models (released by OpenAI) \cite{brown2020language}, the 
LLaMA 
model (released by Meta), and the GLM130B model (released by Tsinghua University). However, due to the strong commercial promotion and good product design provided by Microsoft and OpenAI, the ChatGPT application built on top of the GPT3.5/4 series of models is being increasingly adopted and used by researchers and enterprises in various real-world scenarios, such as intelligent customer service \cite{george2023review}, interactive translation \cite{lu2023error}, and personal assistants \cite{shafeeg2023voice}. Considering the cost of experimentation, this paper's LLMs are based on OpenAI's GPT3.5 and use gpt-3.5-turbo-0301 as the kernel model, using application programming interfaces (APIs) provided by OpenAI.

\subsection*{Personalized Book Recommendations}

Book understanding and recommendation is a fundamental application problem in the field of LIS. With the continuous increase in the number of book resources, both the number of book types and the number of interactions with readers are rapidly increasing. Therefore, how to select suitable books from a massive candidate set for recommendation is a fundamental problem. Generally, we can use personalized recommendation models to solve this typical information overload problem. In existing recommendation systems, the basic definition of a user's preference probability for an item $i$ can be represented by the following function:
$$y_{i \to u} = f(h_i, h_u) \in [0,1]$$
where $h_i$ and $h_u$ represent the learned item feature representation and user feature representation, respectively, and $f(\cdot)$ represents the scoring function that matches the user and item features, such as the cosine similarity function and multilayer perceptron module. Therefore, the existing research on personalized book recommendation can be partly summarized as follows.

\textbf{Collaborative Filtering (CF)-Based Methods}: CF is a classic and practical book recommendation algorithm based on the similarity between readers or books. The basic idea is that if user 1 likes book A and user 2 also likes books A and B, it can be assumed that there is a certain similarity between user 1 and user 2. Therefore, the other items liked by user 1 can be recommended to user 2, or vice versa. CF models can be divided into two types: user-based CF\cite{tewari2014book,bellogin2014neighbor} and item-based CF models\cite{linden2003amazon,rajpurkar2015book}. CF models usually require a large amount of user behaviour data for training and prediction, so they are difficult to use in cases with sparse data or cold-start situations.

\textbf{Deep Learning (DL)-Based Methods}: In recent years, an increasing number of DL-based book recommendation algorithms have been proposed to better address the issues affecting CF models. DL models can learn more complex feature representations from user interaction data, thereby improving their recommendation accuracy. Specifically, DL-based book recommendation algorithms can be divided into two types: matrix factorization-based models \cite{xu2018novel,dien2022approach} and sequence-based models \cite{lipton2015critical}.

\textbf{Graph Neural Network (GNN)-Based Methods}: A GNN-based recommendation system is a recently proposed algorithm that combines graph theory and DL\cite{wu2022graph}. A GNN-based book recommendation system represents readers, books, and their interaction information as a graph and then uses the GNN model to learn and aggregate the node and edge features in the graph, obtaining higher-order feature representations for different recommendation tasks\cite{ma2020memory,wang2021learning}.

\section*{Methods}

\subsection*{Overview}

This paper proposes a BookGPT. By combining the existing LLMs with typical tasks found in book recommendation scenarios, this framework constructs appropriate prompt strategies based on different task features and combines data validation, backtracking, and retrying methods to explore the possibility of using LLMs in book recommendation scenarios. The BookGPT framework is divided into four modules: (1) book recommendation task definition and data preparation, (2) prompt engineering, (3) GPT-based interaction and response parsing, and (4) task evaluation.

 \begin{figure}[H]
	\centering
	\includegraphics[width=\linewidth]{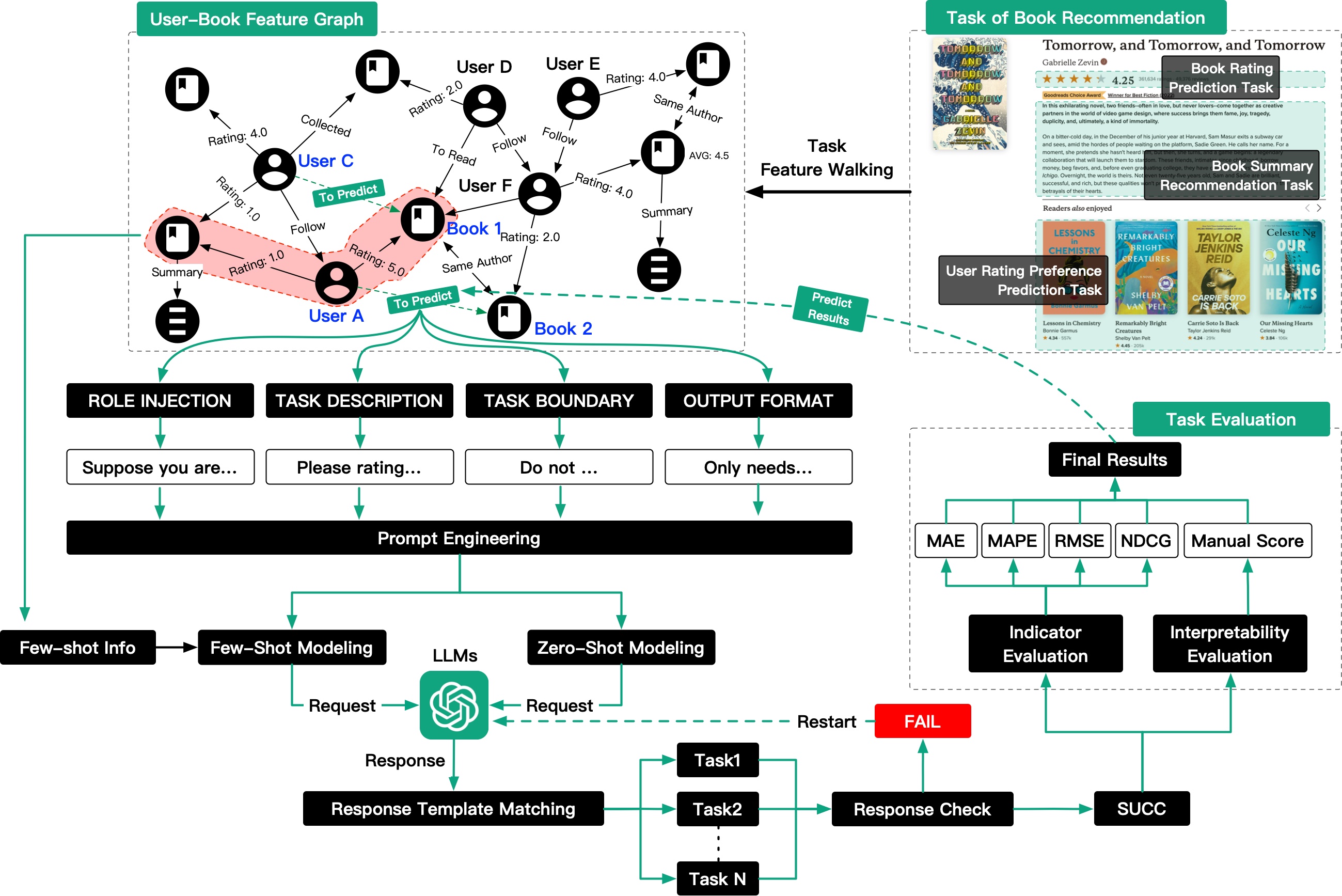}
	\caption{\label{model_framework} Framework of the BookGPT.}
\end{figure}

\textbf{Book recommendation task definition and data preparation.} The BookGPT framework is mainly targeted at three typical book recommendation applications, including book rating recommendation, user rating recommendation, and book content summary recommendation. These three scenarios correspond to the following aspects in the book business: high-quality new book screening based on rating potential, personalized recommendation based on reader preferences, and book explainability recommendation based on content summaries. In these three scenarios, the feature system can be divided into three categories: basic user attribute features, book resource attribute features, and user-book interaction behaviour features. Therefore, it is usually possible to construct recommendation strategies for zero-shot and few-shot settings based on different application scenarios and data enrichment situations to improve the satisfaction achieved by the resulting recommendations. The specific formalizations and definitions of the above tasks are described in detail in Section "Book Recommendation Task Definition".

{\textbf{Prompt engineering.}} Unlike traditional recommendation systems, the core recommendation module of the BookGPT is composed of an LLM, its recommendation process depends on the model's understanding and representation of natural language commands, and the output results are also highly flexible. Therefore, by designing appropriate prompt formats, the model's ability to understand tasks and the effectiveness of the final output results can be effectively enhanced \cite{liu2022design}. Basic prompt engineering involves four core parts: an injected identity, a task description prompt, a task boundary prompt, and an output format prompt. At the same time, the prompt format of CoT \cite{zhao2023survey} can be used to guide the model to solve complex tasks in a step-by-step manner and increase the modelling accuracy of the model. For a detailed analysis of the prompt engineering process of the BookGPT models, please refer to Section "Prompt Engineering for Book Recommendation".

\textbf{GPT-based interaction and response parsing.} The BookGPT model proposed in this article is tested and modelled using the ChatGPT API provided by OpenAI\footnote{https://platform.openai.com/docs/api-reference}. The LLM used is the "gpt-3.5-turbo-0301" model, which is an optimized version based on GPT-3.5. It has faster response times and is approximately 10 times less expensive to call than the base model of GPT-3.5 with the same number of token requests. At the same time, due to the addition of many random factors in the response process of ChatGPT (to prevent the returned answers from being too convergent), it is necessary to perform a targeted formal verification for each returned result. If the obtained response does not satisfy the requirements of the established task, it is necessary to improve the prompt method or try to request it again. For a detailed description, please refer to Section "Output Verification and Task Restarting".

\textbf{Task evaluation.} As the BookGPT is the first framework to apply an LLM to the book recommendation system scenario, no evaluations of the application efficiency and feasibility of this scenario have been performed in previous studies. Therefore, we attempt to design two evaluation strategies, including a metric evaluation and an interpretability evaluation, to evaluate and discuss the performance achieved by the BookGPT on the three key tasks in the book recommendation scenario. Through detailed empirical research, we explore and analyse the advantages and problems of the BookGPT in the book recommendation scenario and conduct relevant discussions on subsequent research directions. The detailed analysis of this part can be found in Sections "Task Evaluation" and "Experiments".

\subsection*{Book Recommendation Task Definition}

\subsubsection*{Book Rating Recommendation}

The book rating task is one of the fundamental tasks in book recommendation scenarios, especially those such as introducing new books and performing book evaluations. To effectively evaluate the ability of the recommendation system constructed by ChatGPT-like LLMs on this task, this paper verifies it through two modelling methods, zero-shot modelling and few-shot modelling, and measures the quality of the recommendation system by examining the difference between the rating results of the discriminant system and the actual rating results. The specific task definition is as follows.

{\textbf{Zero-Shot Modelling.}} Given a book name $b_x$ and an author name without any background information, the system is required to output a rating result $R_{b_x} \in [0,10]$ for the corresponding book, where a higher score indicates that the book is more recommended for reading.

{\textbf{Few-Shot Modelling.}} Compared to zero-shot testing, the essence of few-shot testing is to enhance the model's understanding of the given task by providing partial sample information, with the hope of improving the final prediction performance. Therefore, in this paper, the few-shot testing case for book rating is defined as follows: given a list of books with their corresponding ratings as pairs for the same author, $P_{u_i} = {(b_1,R_{b_1}),(b_2,R_{b_2}),...,(b_n,R_{b_n})}$, a small portion of them (e.g., $k$) are selected as known input information, and the system is required to rate the remaining books from the same author. Finally, the system's rating results are evaluated by the difference between the predicted ratings and the actual ratings of the remaining samples.

\subsubsection*{User Rating Preference Recommendation}

The user rating recommendation task has a wider range of application scenarios than the book rating task, such as predicting the book preferences of users for e-commerce sales, predicting the interest levels of readers in book borrowing, predicting clicks, and predicting library browsing. This task typically uses historical interactions (clicks, browsing, borrowing, collecting, commenting, rating, etc.) between readers and books as feature data sources, combines them with basic user attributes and book attributes, and utilizes various machine learning models to build accurate recommendations. In this article, the specific task is defined as follows.

{\textbf{One-Shot User Preference Modelling.}} Given a historical book behaviour sample sequence (such as a rating sequence) for user $u_i$, $H_{u_i} = \{b_1,b_2,...,b_n\}$, the model is only provided with a single training sample as a hint or training set and is required to score the remaining samples in the behaviour sequence. The final evaluation is based on the consistency between the model's scoring results and the original sample results.

{\textbf{N-Shot User Preference Modelling.}} Given a historical book behaviour sequence (such as a rating sequence) for user $u_i$, $H_{u_i} = \{b_1,b_2,...,b_n\}$, a certain proportion of the data is selected as the training set (or prompt set) from it. The model is required to score the remaining sequence based on the provided training set, and the final evaluation is the consistency between the model's scoring results and the original sample results.

\subsubsection*{Book Summary Recommendation}

The task of book summary recommendation aims to automatically extract concise and accurate summary content from books, providing readers with a quick way to understand the main contents of the books. This task typically utilizes NLP techniques, including text summarization, text classification, information extraction, and other techniques, to achieve its goals. In practical applications, summaries can serve as important data sources for book recommendation, search result previewing, knowledge graph construction, and other areas. Therefore, in this section, we compare the summaries generated by the ChatGPT model with the standard summaries produced by humans and evaluate the effectiveness of the proposed recommendation system from the perspectives of interpretability and credibility. We answer two questions. (1) Can large-scale language models such as ChatGPT and WenXinYiYan (WenXin)\cite{wenxin2023} (released by Baidu) achieve better results than humans in book summary generation tasks? (2) Will ChatGPT and Wenxin exhibit different summarization abilities for different categories of literary genres, such as novels, essays, and poetry? The specific generation forms of the comparison task include the following.

{\textbf{Summaries Without Length Limitations.}} This task generates summary recommendations based on specified author and book title information, with no limit imposed on the character length.

{\textbf{Summaries with Length Limitations.}} 
Because the summaries generated by LLMs usually contain more characters than those of humans, to further ensure the fairness of the comparison, the maximum number of characters that can be generated is further limited when generating an abstract with the model. Specifically, the model is required to generate summary recommendations based on the specified author and book title information, with a forced limit imposed on the maximum number of characters that can be generated; this limit is the same as the character count of the manual abstract provided for the same book.

\subsection*{Prompt Engineering for Book Recommendation}

\begin{figure}[H]
	\centering
    \includegraphics[width=.8\linewidth]{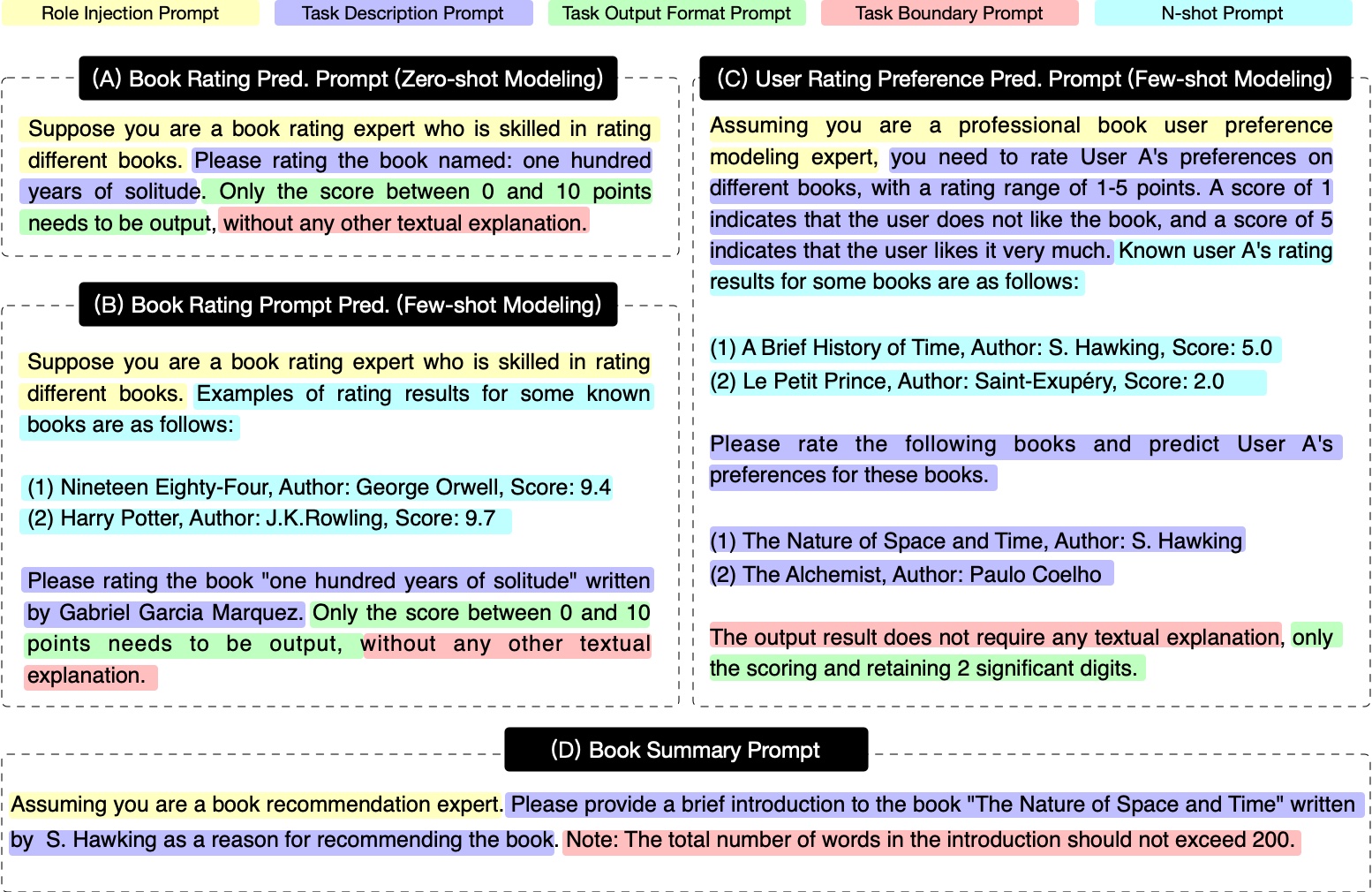}
            \vspace{-0.2cm}
    \caption{Prompt examples for the BookGPT.}
    \label{prompt_example_all}
\end{figure}

As far as we know, LLMs are typical generative language models, so the quality of the output contents of these models usually exhibit significant correlations with the input prompt contents. Therefore, in this section, we discuss how to design effective prompt content\cite{PromptGuide2022} for various types of book recommendation tasks to achieve improved recommendation efficiency. As shown in Figure \ref{prompt_example_all}, we provide prompt engineering examples for three typical book recommendation scenarios. Generally, prompt content typically includes four parts.

(1) {\textbf{Role Injection Prompt.}} This prompt is mainly used to indicate the role type represented by the LLM, guiding it to respond differently according to specific role types. For example, in the first task of book recommendation, the book rating task, if no prompt is given for identity injection ("Assuming you are a professional book rating expert") and ChatGPT is instead directly asked to answer a task requirement such as "Please rate the book xxx", ChatGPT usually responds with a refusal to answer.

 \begin{figure}[H]
	\centering
	\includegraphics[width=0.6\linewidth]{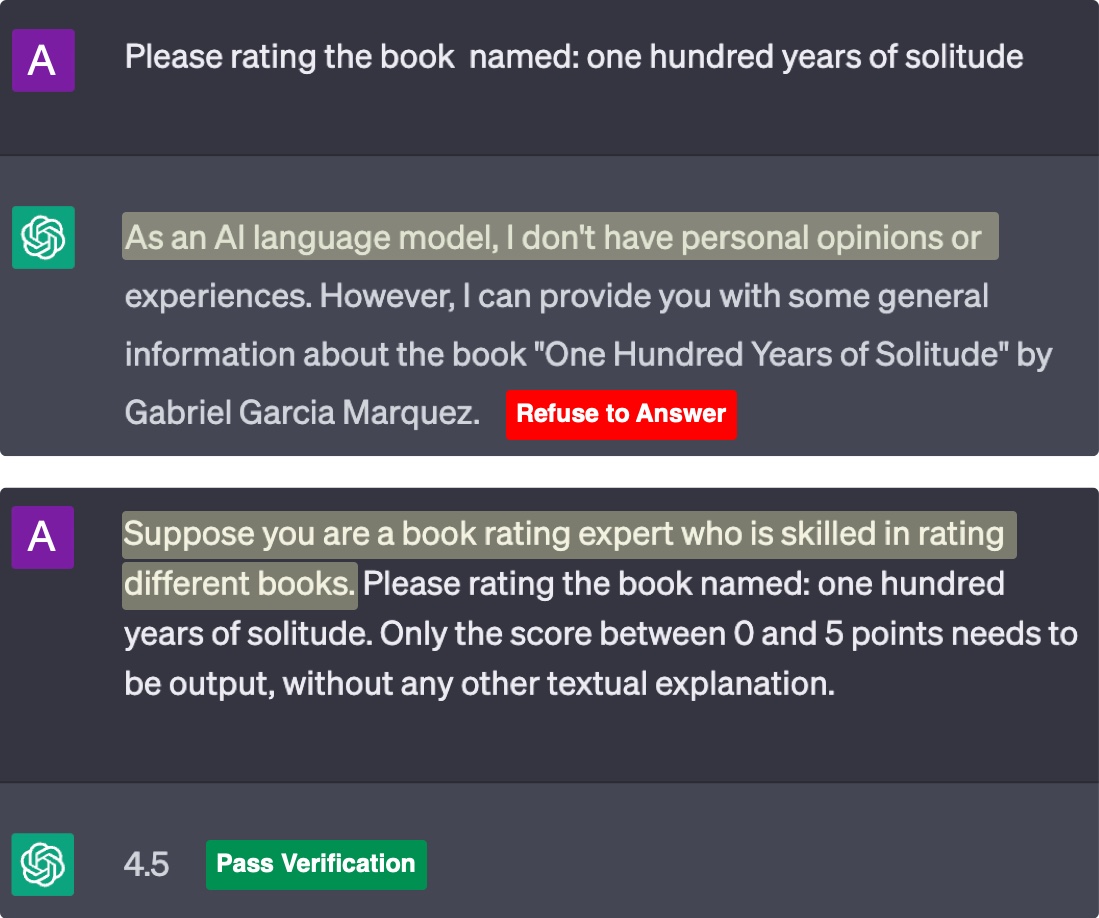}
        \vspace{-0.2cm}
	\caption{Example of role injection.}
    \label{role_like}
\end{figure}
 
(1) {\textbf{Task Description Prompt.}} This prompt is used to provide information about the specific task that ChatGPT is being asked to perform. It typically includes details such as the type of task, the goal of the task, and any relevant information about the input data or context. This prompt helps ChatGPT understand the task and generate more accurate and relevant responses. For example, in a book recommendation task, the task description prompt might include information about the user's reading preferences, the desired book genre, and any specific requirements or constraints for the recommendation process.

Certainly, due to the current limitations of model inference capabilities, the lengths of task description prompts are usually restricted within a certain range (the current GPT3.5 version limits the total number of tokens for model responses and prompt contents to 4096\footnote{https://platform.openai.com/tokenizer}) to obtain better model inference results. Typically, content prompts based on task examples (few-shot scenarios) serve as additional training samples for the model, which can further enhance the model's fitting and understanding of the task, ultimately leading to better prediction results.

(3) {\textbf{Task Boundary Prompt.}} This prompt is used to impose negative limitations on ChatGPT-like models, telling them what not to do in a given task. For example, in a book rating task, if only identity injection and task description prompts are used, the model will often generate the corresponding rating and a long explanatory passage, which can cause difficulties for downstream applications. Therefore, it is necessary to explicitly provide limits and tell the model what the task boundary is; that is, no textual explanation is needed, and only the rating result should be output. At this point, the model will only produce the corresponding rating score as needed.

(4){\textbf{Task Output Format Prompt.}} 
After completing the role injection, task description, and boundary prompts in the BookGPT, we also need to tell the model about the final output format. The main benefits of this prompt are (1) improving the accuracy of the model's output results: by specifying the output format of the model, we can ensure that the results produced by the model satisfy the requirements of downstream systems, avoiding errors and unnecessary additional processing steps caused by mismatched data formats; And (2) enhancing the maintainability of the system: by explicitly specifying the output format of the model, we can avoid the need for downstream systems to make significant modifications and adjustments when the model's output format changes, thereby enhancing the system's maintainability and scalability. For example, for the book rating prediction task, the output format needs to be limited to a value with two decimal places. For the user preference estimation task, the output format needs to be limited to a Python list format.

\subsection*{Output Verification and Task Restarting}

Through the design of a prompt engineering strategy, we can ensure that the output of the model satisfies the expected definition to some extent. However, since ChatGPT-type models are essentially natural language probability models, and because ChatGPT incorporates stochastic factors into its design to ensure the diversity of the generated results \cite{zhao2023survey}, it is possible that the model may produce different response results for the same input request. Therefore, in the end, we also need to recheck the legality of the generated content produced by ChatGPT-type models, that is, perform a secondary verification on the critical output data format and the requirements.

In this module, we build independent validation functions for each type of book recommendation subtask. For example, for the book rating task, we need to check if the returned result is a numerical value. For the user rating recommendation task, we need to extract the rating values corresponding to the books in the returned result and check if the number of returned results matches the requested quantity. For the book summary task, we need to check if the length of the returned text satisfies the input requirements. If the result returned by ChatGPT does not meet the specified format requirements, we need to resend the request to the recommendation module until the maximum number of retries is reached (the maximum number of retries in this framework is set to 3).

\subsection*{Task Evaluation}

To validate the performance of the BookGPT (including zero-shot and few-shot modelling), we evaluate the system from two aspects: a task metric evaluation and an interpretability evaluation.

{\textbf{Task metric evaluation:} We evaluate the recommendation performance of the BookGPT on two subtasks: book rating recommendation and user rating preference recommendation. Specifically, we treat the book rating task as a regression model and evaluate the performance of different recommendation models in terms of the mean absolute error (MAE), mean absolute percentage error (MAPE), and root mean square error (RMSE) metrics. For the user rating preference recommendation task, which is treated as a sorted rating recommendation process, in addition to focusing on the MAE, MAPE, and RMSE, we further evaluate the performance of different models in terms of the normalized discounted cumulative gain (NDCG) metric. The specific calculation methods for each metric are shown in Table \ref{eva_index_cal}.

\begin{table}[H]
 
	\centering
	\caption{Task evaluation metrics. }
    \label{eva_index_cal}  
    
	\begin{tabular}{lccc}
		\toprule
		Metrics     & Equations  & Concerns   \\
		\midrule
		  MAE     & $\frac{1}{n} \sum^{n}_{i=1} \left|{y}_{i} - \tilde{y}_{i} \right|$   & absolute error   \\
		MAPE      & $\frac{1}{n} \sum^{n}_{i=1} \left| {y}_{i} - \tilde{y}_{i} \right| / \left| {y}_{i} \right| $   & percentage error    \\
		  RMSE & $\sqrt{ \frac{1}{n}  \sum^{n}_{i=1} \left( {y}_{i} - \tilde{y}_{i} \right)^2 } $ & divergence   \\
            NDCG@$k$ & Details\cite{wang2013theoretical} & cumulative gain   \\
		\bottomrule
	\end{tabular}
\end{table}

{\textbf{Interpretability Evaluation:}} This part mainly evaluates the recommendation ability of the BookGPT in the book summary recommendation task. The goal of this task is to identify the key content of a book through its summary and arouse readers' interest in reading or purchasing it. Therefore, in this evaluation, we use manually generated summaries as the ground truth to evaluate the book summary generation performance of the BookGPT model when different testing books. Specifically, since ChatGPT was trained on English corpora, we also introduce a large Chinese language model, Wenxin (released by Baidu Group) as a reference to test the performance of an LLM on Chinese summaries and compare it with ChatGPT. Finally, through the above two types of evaluations and empirical studies, we attempt to answer the following three questions.

\begin{itemize}
    \item Q1: What tasks in the book recommendation scenario are the BookGPT suitable for? How is its performance?
    \item Q2: Is there a significant difference between the final recommendation performances achieved with zero-shot modelling and few-shot modelling in the BookGPT?
    \item Q3: In the book recommendation scenario, what are the potential research directions concerning ChatGPT-like LLMs in the future? What problems can they solve?
\end{itemize}

\section*{Experiments}
 
As shown in Table \ref{datasets_info}, the experimental datasets in this article include three types of data: book rating data, book summary data, and book-user interaction data.

\begin{table}[H]
	\centering
	\caption{Datasets.}
    \label{datasets_info}
	\begin{tabular}{lccc}
		\toprule
		Datasets     & Types  & Amount   \\
		\midrule
		Douban Rating     & scores   & 3228 books   \\
		Douban Book Summary      & summary text   & 50 books    \\
		Goodbook-10k  & scores, interactions & 10k books   \\
		\bottomrule
	\end{tabular}
\end{table}

{\textbf{Douban Rating:}} This dataset was collected from the book rating channel of Douban\footnote{https://book.douban.com/}, which includes four key fields: book title, author, rating, and number of comments. Due to the limited number of API requests for OpenAI, only books with more than 2000 ratings are selected as the test dataset, resulting in 3228 popular books for evaluation purposes. In this experiment, we evaluate the performance of the BookGPT under zero-shot learning and different levels of few-shot learning (1/3-shot learning and 2/3-shot learning), and the core observation indicators are the MAE, MAPE, and RMSE.

{\textbf{Douban Book Summary:}}To evaluate the model's ability to summarize and recommend book content, we select 50 popular books from the Douban TOP250 book channel\footnote{https://book.douban.com/top250}, including 4 categories of literary genres and their human-written summaries: 20 novels, 10 essays, 10 poems, and 10 dramas. The human-written summaries are provided by Douban's book editing experts and are used as the benchmark for the comparison. The summary results of the BookGPT model and the Wenxin model in the "restricted size" and "free size" summary scenarios are used as comparison models.

Notably, to ensure the fairness of the evaluation results, we first randomly mix the human-written summaries and model summaries and present them in a random order to different annotators. Then, 15 annotation participants are asked to rank the summaries, and each summary must be annotated for at least 3 minutes. If a summary is ranked higher, it is considered to have a stronger recommendation ability and more attractiveness. Finally, the results acquired from different annotators are processed to obtain the evaluation results for each model with respect to different literary genres of books and overall. The core observation indicators for this task are the summary evaluation score and average summary length. The calculation method for the final score of each model's summary is as follows:

$$Score_{M_{1}}\  =\  \frac{1}{N} \sum^{rank}_{i} freq\times w_{i}$$

In the above formula, N is the total number of annotators in the test, and $freq$ is the number of times the corresponding model option appears in position $i$ with a weight of $w_i$. For example, assuming that 15 people participate in annotation sorting, three options need to be sorted, with positions 1, 2, and 3 corresponding to scores of 3, 2, and 1, respectively. Then, if model X's sorted summary results for book Y are first place 10 times, second place 2 times, and third place 3 times, the comprehensive score of model X's summary for book Y in this test is (103+22+3*1)/15 = 2.47 points.

{\textbf{GoodBook-10k:}} The GoodBook-10k dataset\cite{goodbooks2017} was collected from the Goodreads\footnote{https://goodreads.com/} book review website, which is the largest online reading community in the world and is similar to Douban Reading in China. The GoodBook-10k dataset contains rating data for 10,000 popular books and 5.98 million users' ratings, with fields including book ratings, user bookshelf labels, book metadata, and book tags. In this paper, we use it for the user rating task, which includes three forms at the prompt level: 1-shot, 10-shot, and 20-shot learning, where the model is provided with 1, 10, or 20 model rating records, respectively, and is required to predict the remaining records and provide user preference rankings. The benchmark models for this task include the BookGPT model proposed in this article, as well as four classic CF-based recommendation algorithm models for personalized recommendation scenarios: the matrix factorization model (FunkSVD)\cite{mnih2007probabilistic}, the K-nearest neighbours (KNN; means) model\cite{koren2010knnmeans}, the SlopeOne\cite{lemire2005Slope} model, and the CoClustering\cite{george2005CoClustering} model. The evaluation metrics are the NDCG@{5,10,15,20}, MAE, MAPE, and RMSE.

\section*{Results}

This section analyses the performance achieved by the BookGPT model and the baseline models on different tasks, answering the questions raised above; namely, how does the BookGPT model perform on different tasks in the book recommendation scenario? Can few-shot learning improve the recommendation performance of the BookGPT model?

{\textbf{Book Rating Task.}} As shown in Table \ref{exp_res_book_rate}, the prediction results yielded by the BookGPT model in book rating tasks are analysed under zero-shot modelling, 1/3-shot modelling, and 2/3-shot modelling. Overall, the BookGPT model exhibits a good regression prediction ability, with MAPE values ranging from 8.8\% to 5.4\%, indicating that it performs well on book rating tasks. The overall absolute percentage error is within a small range of 10\%, and even for 2/3-shot modelling, the MAPE can reach an estimated value of 5.4\%.

\begin{table}[H]
	\centering
	\caption{Results of the book rating task.}
    \label{exp_res_book_rate}  
	\begin{tabular}{lllll}
		\toprule
		BookGPT   & MAE   & MAPE & RMSE  \\
		\midrule
		Zero-shot modelling    & 0.682    & 0.088  & 0.886   \\
		1/3-shot modelling  &  0.441 \textcolor{mygreen}{($\downarrow$35.34\%)}   & 0.057 \textcolor{mygreen}{($\downarrow$35.23\%)}  & 0.558 \textcolor{mygreen}{($\downarrow$37.02\%)}   \\
		2/3-shot modelling  & \emph{0.419} \textcolor{mygreen}{($\downarrow$38.56\%)}   & \emph{0.054} \textcolor{mygreen}{($\downarrow$38.64\%)}  & \emph{0.538} \textcolor{mygreen}{($\downarrow$39.28\%)}    \\
		\bottomrule
	\end{tabular}
\end{table}

Furthermore, based on the results of few-shot modelling with prompt enhancement, the accuracy of the estimated rating is significantly improved compared to that of zero-shot modelling. The model's MAE value decreases from 0.682 (zero-shot) to 0.441 (1/3-shot modelling) and 0.419 (2/3-shot modelling), representing 35.34\% and 38.56\% mean absolute error decreases compared to that of zero-shot modelling, respectively. In addition, the model's RMSE also decreases significantly, from 0.886 to 0.558 (1/3-shot modelling) and 0.538 (2/3-shot modelling), with a relative decrease in the optimal mean square error of 39.28\% (2/3-shot modelling). This result indicates that {\textbf{few-shot modelling can significantly reduce the prediction error induced by zero-shot modelling in the book rating task by providing reference samples for the BookGPT model.}}

Finally, comparing the results of the BookGPT model with those of different levels of prompt-based few-shot learning, increasing the prompt size again (from 1/3 to 2/3 of the training set) leads to an additional improvement in the final performance. However, the improvement from zero-shot learning to 1/3-shot learning is larger than the improvement from 1/3- to 2/3-shot learning. This is because the 1/3 prompt size already provides a good information reference for the BookGPT model, and further increasing the prompt size brings limited information gain while increasing the model's reasoning overhead. Therefore, in practical applications, the appropriate prompt size can be selected based on the information gain inflection point through ablation experiments and in combination with the scenario's needs, thus effectively yielding improved few-shot learning performance.

{\textbf{User Rating Preference Recommendation Task.} Figure \ref{user_rate_all_figure_ndcg} represents the NDCG evaluation results obtained for the user rating preference recommendation task.

\begin{figure}[H]
\centering
\subfigure[1-shot, NCDG@k=\{5,10,15,20\}]{\label{fig:subfig:a}
\includegraphics[width=0.328\linewidth]{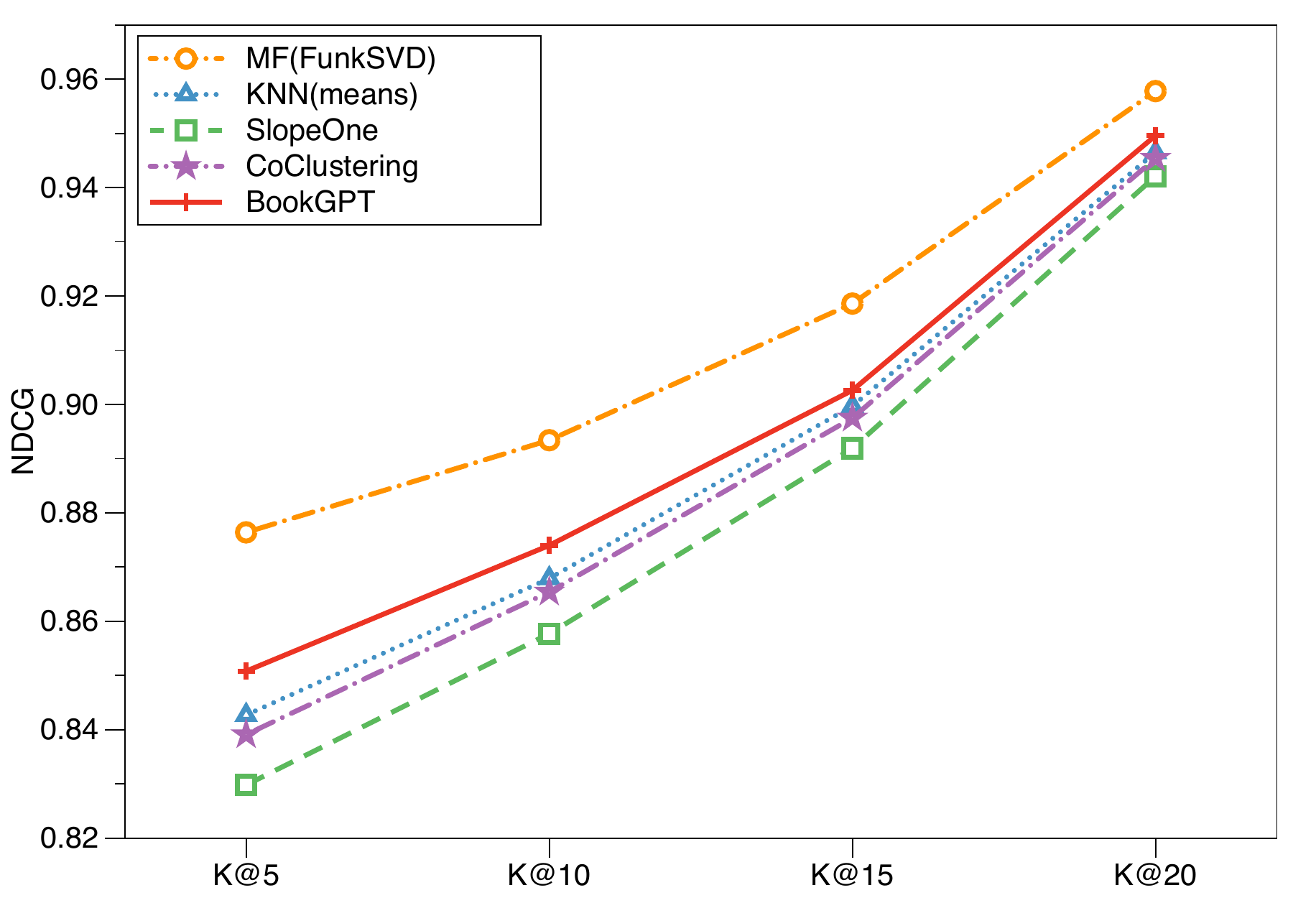}}
\hspace{-0.01\linewidth}
\subfigure[10-shot, NCDG@k=\{5,10,15,20\}]{\label{fig:subfig:b}
\includegraphics[width=0.328\linewidth]{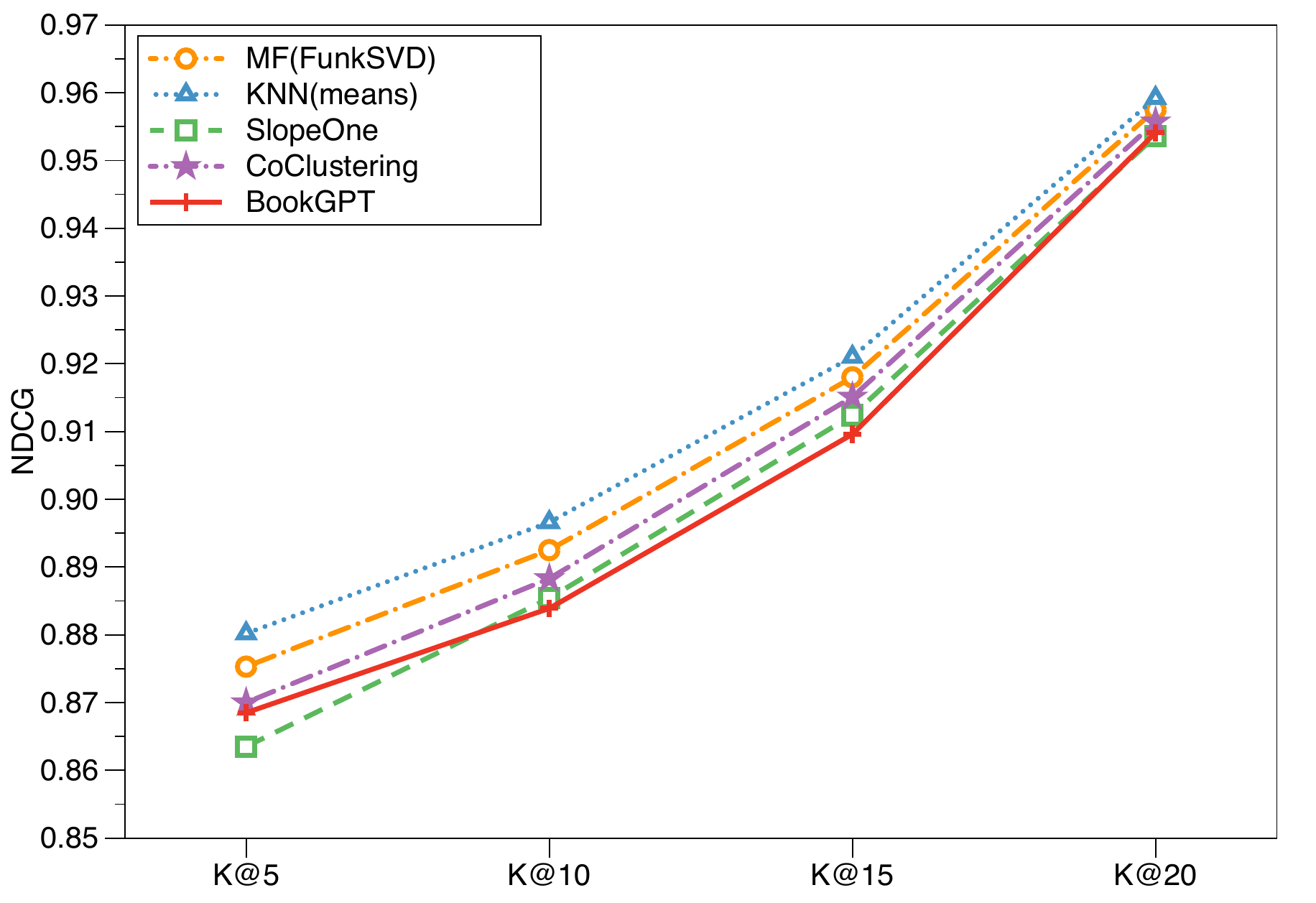}}
\hspace{-0.01\linewidth}
\subfigure[20-shot, NCDG@k=\{5,10,15,20\}]{\label{fig:subfig:c}
\includegraphics[width=0.328\linewidth]{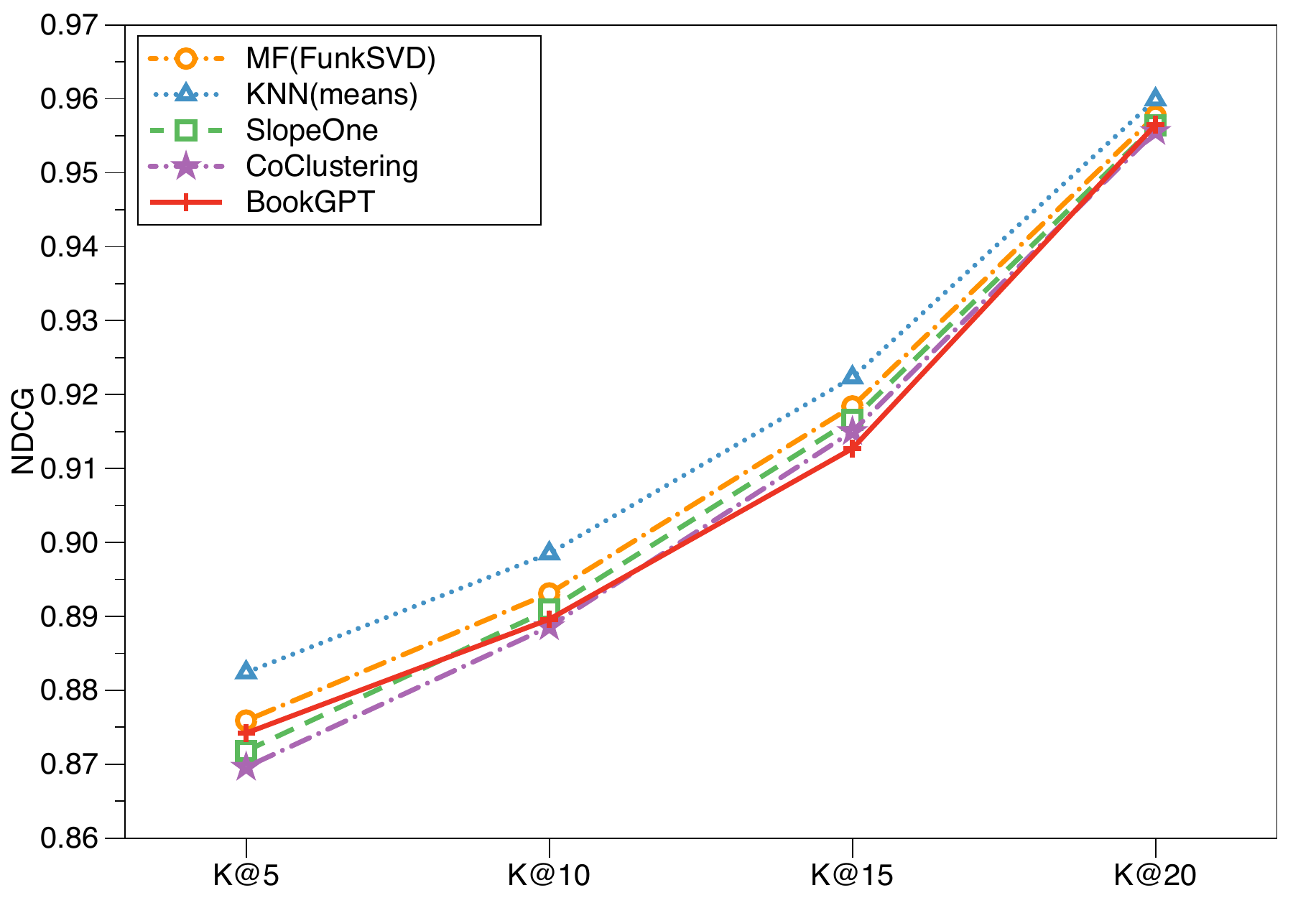}}

\vfill
\subfigure[NDCG@5, prompt items = \{1,10,20\}]{\label{fig:subfig:d}
\includegraphics[width=0.328\linewidth]{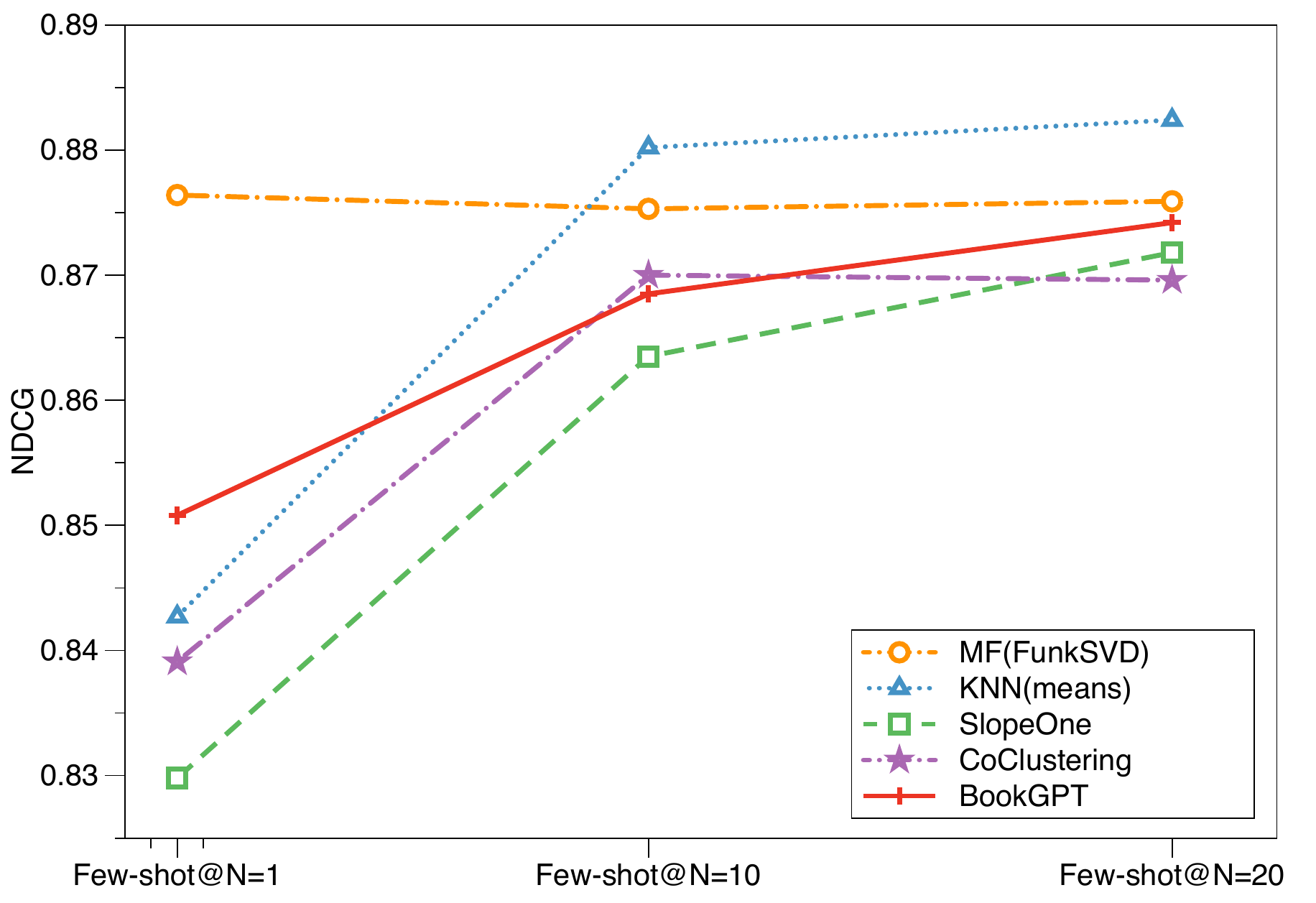}}
\subfigure[NDCG@20, prompt items = \{1,10,20\}]{\label{fig:subfig:e}
\includegraphics[width=0.328\linewidth]{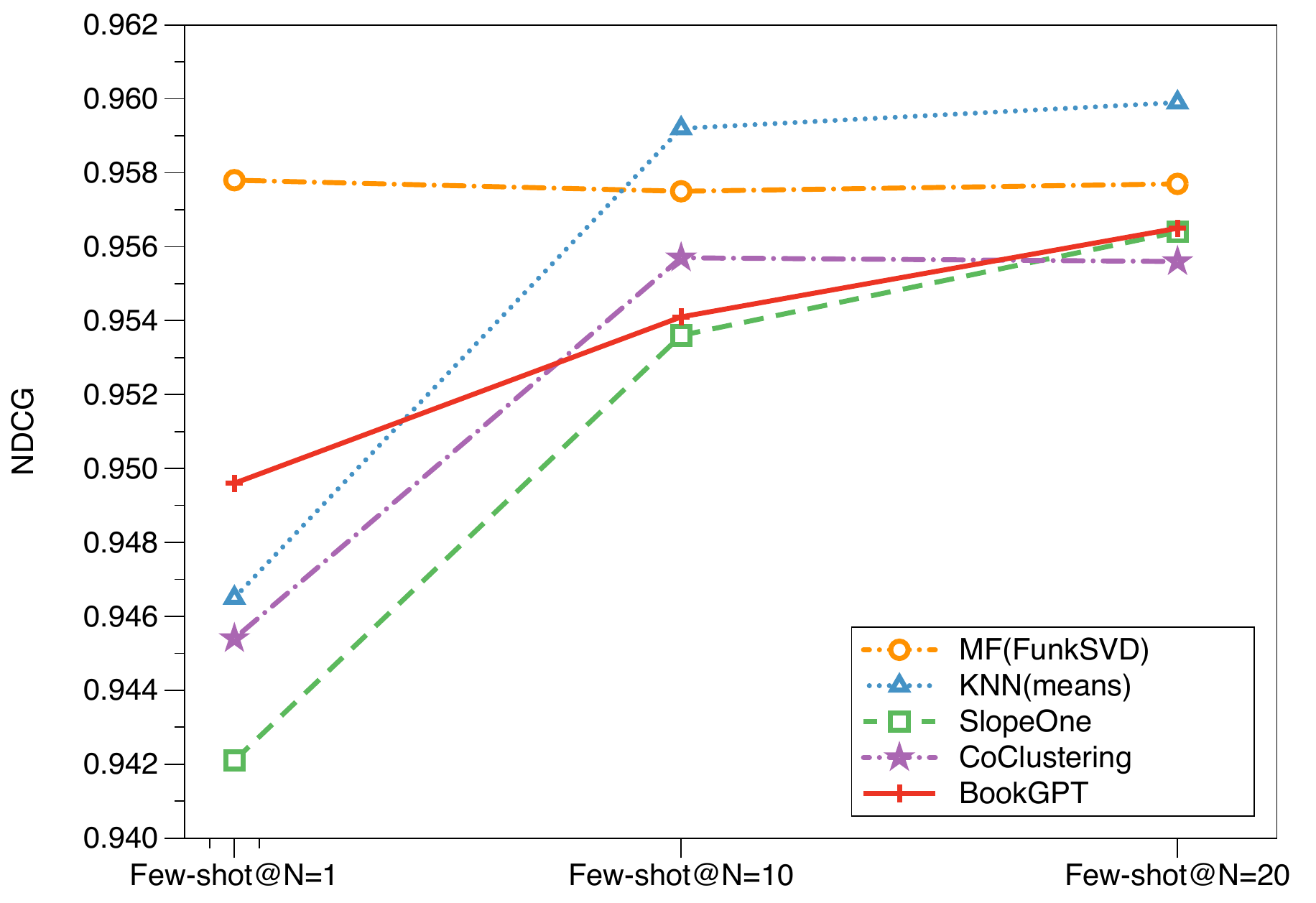}}
\label{fig:subfig}

\caption{NDCG scores obtained in the user rating preference recommendation task.}
\label{user_rate_all_figure_ndcg}
\end{figure}

First, overall, the MF (FunkSVD) model can achieve good results in different subtasks, especially in the single-sample scenario, where it performs best. The reason for this is that the recommendation strategy based on FunkSVD models the user-book rating interaction matrix through matrix factorization. The optimization goal of this modelling method is to make the residual between the user ratings and the rating product obtained by matrix multiplication as small as possible. Therefore, even with limited reference rating information provided by the user to be predicted (such as in 1-shot learning), FunkSVD can still achieve good results in terms of metrics such as the RMSE. However, as the number of effective prompt samples (features) for a single user increases, clustering models represented by KNN (means) begin to perform better. During the prediction process, KNN (means) relies on modelling the user's historical rating habits to generate the final estimation result, leading to a prediction accuracy increase with the increase in the number of prompt samples. Notably, the BookGPT also achieves the second-best recommendation results in terms of the NDCG metric in the one-shot scenario, which once again confirms the advantage of the BookGPT in small sample recommendation scenarios. However, it is also observed that {\textbf{compared with the application of classic recommendation models, the application effect of the LLM-based BookGPT still has worse personalized user understanding and modelling.}}

\begin{table}[H]
\centering
\caption{The prediction errors induced in the user rating preference recommendation task.}
\label{eva_user_rate} 
{
\begin{tabular}{cl|ccc}
\toprule
\multirow{2}{*}{Prompt items}        & \multirow{2}{*}{Models}  & \multicolumn{3}{c}{Indicators}                                      \\
 &
   &
  \multicolumn{1}{c}{\textbf{MAE}} & 
  \multicolumn{1}{c}{\textbf{MAPE}} &
  \multicolumn{1}{c}{\textbf{RMSE}} \\
  \midrule
\multirow{5}{*}{1-shot modelling}  & FunkSVD\cite{mnih2007probabilistic}             & \colorbox{gray!30}{0.765} & \colorbox{gray!30}{0.259} & \colorbox{gray!30}{0.956} \\
                             & KNN (means)\cite{koren2010knnmeans}       & 0.865 & 0.283 & 1.153 \\
                             & SlopeOne\cite{lemire2005Slope}           & 0.843 & 0.278 & 1.117       \\
                             & CoClustering\cite{george2005CoClustering}       & \emph{0.836} & \emph{0.276} & \emph{1.107}      \\
                             & BookGPT                &   1.075	&   0.297	&   1.342     \\\midrule
                    
\multirow{5}{*}{10-shot modelling} & FunkSVD\cite{mnih2007probabilistic}           & \emph{0.733} & 0.249 & \emph{0.917} \\          
                             & KNN (means)\cite{koren2010knnmeans}   &  \colorbox{gray!30}{0.708}  &  \colorbox{gray!30}{0.236}  &  \colorbox{gray!30}{0.912}      \\
                             & SlopeOne\cite{lemire2005Slope}            & 0.741 & \emph{0.242} & 0.965       \\
                             & CoClustering\cite{george2005CoClustering}        & 0.737 & 0.243 & 0.961      \\
                             & BookGPT                	& 0.915	& 0.263	& 1.184     \\\midrule
\multirow{5}{*}{20-shot modelling} & FunkSVD\cite{mnih2007probabilistic}             & 0.710 & 0.241 & \emph{0.892} \\
                             & KNN (means)\cite{koren2010knnmeans}   &   \colorbox{gray!30}{0.676}  &   \colorbox{gray!30}{0.226}  &  \colorbox{gray!30}{0.879}      \\
                             & SlopeOne\cite{lemire2005Slope}         & \emph{0.700} & \emph{0.233} & 0.905        \\
                             & CoClustering\cite{george2005CoClustering}          & 0.703 & 0.235 & 0.910     \\
                             & BookGPT              	&0.842	&0.248	&1.116    \\
\bottomrule
\end{tabular}
} \end{table}

Second, comparing the results based obtained on different prompt sample quantities, the NDCG score of the FunkSVD model is not sensitive to the number of samples, and its performance remains consistent across the 1-shot, 10-shot, and 20-shot subtasks. However, other baseline models, such as KNN (means), SlopeOne, CoClustering, and the BookGPT model, exhibit significant changes in their NDCG scores as the number of prompt samples increases. This is because these models build a model of the user's historical ratings during the result prediction process. For example, in the BookGPT model, during the rating process, the model uses the user's historical book rating preferences as background knowledge to model the user's interest preferences and makes comprehensive predictions for the newly predicted samples by referencing this knowledge. With an appropriate number of prompt samples, the BookGPT model can typically learn real-time contextual knowledge and apply it to the prediction scenario. This capability is also known as in-context learning \cite{zhao2023survey}, which is one of the important foundational capabilities of LLMs. As shown in Table \ref{eva_user_rate}, by increasing the number of prompt samples, the BookGPT model achieves significant improvements in terms of its MAE, MAPE, and RMSE metrics in the 20-shot subtask, with error reductions of 21.67\%, 16.53\%, and 16.84\%, respectively.

Finally, when evaluating different types of metrics, it can be observed that in situations where the prompt samples are relatively few, such as in the 1-shot scenario, the BookGPT model performs well in terms of its ranking ability (NDCG), but it does not show any advantage in terms of error measurement (the MAE/MAPE/RMSE metrics) over the control models. Additionally, increasing the number of prompt samples for a single user from 10 shots to 20 shots does not result in a similar performance improvement in terms of the NDCG, including the BookGPT model. However, the error metric increases are still considerable. Therefore, if the application scenario emphasizes the absolute value preference of each user to be recommended, the effect can be improved by increasing the number of prompt samples for each user. If the focus is only on the relative ranking ability, performing modelling based on a small number of samples can satisfy the imposed requirements and further save inference resources.

{\textbf{Book Summary Recommendation Task.}} During the evaluation process of this task, to ensure the effectiveness of the results, all annotation processes are carried out anonymously with initial order randomization and cross-evaluation, requiring the average ranking results of each model to be obtained across multiple annotators. In addition, because LLMs such as ChatGPT and Wenxin are optimized based on human instructions, people tend to choose longer answers as high-quality answers during the optimization process, so if the summary size of the model is not controlled, the model tends to produce longer content recommendation results. To ensure fairness during the comparison with the human summaries on Douban, we add summary size restrictions to the prompt construction process, requiring that the summary recommendation results of the BookGPT and Wenxin be as close as possible to the number of words in human summaries to ensure the validity of the comparison. Table \ref{book_abstract} shows the results of manual evaluations of the content summary recommendations provided by the ChatGPT-based BookGPT model and Wenxin for different book genres.

\begin{table}[!htbp]
\centering
\caption{The results obtained in the book summary recommendation task.}
\label{book_abstract} 
\begin{threeparttable}
\resizebox{\textwidth}{!}{%
\begin{tabular}{cl|cc|cc|cc|cc|cc}
\toprule
\multirow{3}{*}{Size}    & \multicolumn{1}{c}{\multirow{3}{*}{Models}} & \multicolumn{10}{c}{Genres}                                           \\
 &
  \multicolumn{1}{c}{Models} &
  \multicolumn{2}{c}{All} &
  \multicolumn{2}{c}{Novels} &
  \multicolumn{2}{c}{Poems} &
  \multicolumn{2}{c}{Essays} &
  \multicolumn{2}{c}{Dramas} \\
 
                         & \multicolumn{1}{c}{}                      & Score & Length & Score & Length & Score & Length & Score & Length & Score & Length \\
\midrule
\multirow{3}{*}{Restricted size} & Douban                                    &2.05	&300	&2.07	&323	&2.09	&360	&2.14	&255	&1.88	&237\\
                         &  Wenxin\cite{wenxin2023}                  &\colorbox{gray!30}{2.35}	&251	&\colorbox{gray!30}{2.15}	&278	&\colorbox{gray!30}{2.49}	&291	&\colorbox{gray!30}{2.34}	&204	&\colorbox{gray!30}{2.65}	&206\\
                         & ChatGPT 3.5                               &1.60	&129	&1.79	&178	&1.42	&94	&1.53	&112	&1.47	&84 \\
\midrule
\multirow{2}{*}{Free size}  & ChatGPT 3.5                              &1.45	&314	&1.48	&280	&\colorbox{gray!30}{1.53}	&370	&1.40	&309	&1.39	&329\\
                         & Wenxin\cite{wenxin2023}                   &\colorbox{gray!30}{1.55}	&472	&\colorbox{gray!30}{1.52}	&482	&1.50	&494	&\colorbox{gray!30}{1.60}	&452	&\colorbox{gray!30}{1.61}	&449\\
\bottomrule
\end{tabular}%
}

\end{threeparttable}
\end{table}

First, if we only compare the models based on the limited summary size, Wenxin achieves the best recommendation performance among all the compared models, both in the subgenre and overall tasks. Compared with the human-written summaries from Douban and the ChatGPT model summaries, Wenxin achieves relative improvements of 14.97\% and 47.25\%, respectively.

Furthermore, although we limit and remind the models to pay attention to the number of characters during the prompt construction process, it is apparent that Wenxin and ChatGPT have different understandings of the character limit requirement. During the actual testing process, we find that ChatGPT is more conservative in terms of the character limit rule and usually strictly follows the limit requirement, while Wenxin tends to produce longer summaries. In terms of the average character lengths of the generated summaries, Wenxin exceeds ChatGPT by 94.57\%. Based on this result, we believe that one of the reasons for this is that Wenxin incorporates more Chinese language data into its training and fine-tuning processes and lacks intervention during rule-based prompt fine-tuning, making the model more inclined to produce longer results (with stronger expressions) for Chinese tasks. At the same time, this also indicates that Wenxin has a weaker sense of "rules".

Afterwards, we remove the character limit during the prompt building process, allowing the models to freely generate summary recommendations based on their own capabilities. As shown in Table \ref{book_abstract}, compared to the results obtained with limited character counts, the advantage of Wenxin over ChatGPT is reduced in the free-scale evaluation, with a relative improvement of only 6.89\% compared to ChatGPT, as opposed to 47.25\% under the character limit. This result also suggests that the advantage of Wenxin over ChatGPT under the character limit may be due to the production of longer summaries. However, in terms of actual summary generation ability, Wenxin and ChatGPT are relatively close.

From the performance results obtained for different genres, it can be seen that the improvement exhibited by Wenxin over the Human summaries on the "poems" and "drama" genres is more significant than that on the "novels" and "essays" genres. Furthermore, if the summary sizes of the models are not limited, Wenxin and ChatGPT perform similarly on the "novels" and "poems" genres, while Wenxin's advantage is more obvious on the "essay" and "dramas" genres. However, regardless of the genre, the performance of Wenxin is better than that of the human summaries on Douban.

Overall, in the book content summary recommendation task, the BookGPT based on LLMs has certain advantages over the human-generated summaries on Douban and can provide relatively good improvements for different genres. However, we also discover some issues, such as "fantasizing" and "piecing together" in some summary content. For example, when ChatGPT produces the summary of a book named "Demi-Gods and Semi-Devils", it says, "The Legend of the Condor Heroes is one of Jin Yong's representative works, telling the adventure story in the background of the prosperous Tang Dynasty, and unfolding a multi-linear narrative centered on the protagonist Chen Jialuo with ups and downs." For readers who are not familiar with this book, the result seems reasonable, but in reality, this summary not only describes the wrong dynasty of the story but also uses an incorrect name for the protagonist. In contrast, Wenxin is correct in terms of all key information. Therefore, it can be seen that if one must achieve good fact description results and accuracy in a specific scenario, it is usually necessary to further enhance the training process based on the corpus and prompt rules of that scenario. Otherwise, a model trained on a general language corpus may easily fail to ensure the factual correctness of the generated content.

\section*{Conclusion and Future Work}

This paper proposes a book recommendation framework called the BookGPT based on LLMs, which integrates the understanding and reasoning abilities of LLMs into the classic scenario of book understanding and personalized recommendation in the LIS field. By building a task definition, prompt engineering strategy, interactive querying method, and result verification framework, the article explores the effects of LLM models on three typical subtasks of book recommendation: (1) the book rating task, (2) the user-book preference recommendation task, and (3) the book summary recommendation task.

At the same time, an extensive comparative analysis is conducted based on different prompt sample quantity levels, including zero-shot, one-few shot, and few-shot modelling with 10/20 samples. Finally, from the experimental results, the BookGPT can achieve recommendation effects that are equal to or better than those of classic recommendation models in all three subtasks, especially when the sample size is small, and it has a strong generalization ability. Additionally, as the number of prompt samples increases, the model's recommendation performance significantly improves.

In future work, we will further explore the following research directions concerning the applications of LLMs in LIS and efficiency improvements.

{\textbf{Optimization through task-specific data fine-tuning.}} The current BookGPT framework is built directly on pretrained LLMs such as ChatGPT and Wenxin, and its recommendation performance usually depends on the corpus during LLM training, with a focus on the model's generalization ability. It is not optimized for various proprietary scenarios in the LIS field. Therefore, an important research direction for the future is how to construct training data for fine-tuning on specific domain scenarios to further leverage the knowledge and reasoning advantages of LLMs, improve the recommendation and prediction performance of the corresponding model, and even achieve better performance than that of the current state-of-the-art recommendation models in domain-specific scenarios.

{\textbf{Combining user feedback with multiround conversation-based recommendation.}} In the current BookGPT recommendation paradigm, single-round offline recommendation is adopted, and no attention is paid to user feedback regarding the recommendation effect (such as clicking, pressing the “favourite” button, and borrowing behaviour). Therefore, it is also worth exploring how to integrate different real-time user behaviours and interactions with the system into the recommendation paradigm and construct a multiround conversation-based recommendation model. Through multiround conversation-based recommendation, not only can the contextual learning abilities of LLMs be maximized but also more training prompt language materials can be generated from the interactions, which can improve the model's training and fine-tuning results.

{\textbf{Incorporating personalized user info for explainable recommendations.}} Because LLMs are developed and trained based on various natural language corpora, it is possible to incorporate more personalized user information into the recommendation results and express them in a more "natural" form, rather than merely providing direct recommendations. For example, suppose that a student majoring in history wants to search for books on "recommendation algorithms". If the system can account for the reader's major background attributes when recommending books and explain why a certain book is recommended from a professional perspective, the common points or connections it may have with the reader's major attributes, or what issues need to be noted while reading, could this improve the user's acceptance rate? Therefore, this optimization strategy based on personalized and interpretable recommendations for users is also a very interesting research direction.

In summary, this article aims to explore the possibility of applying LLMs in the LIS field through empirical research and evaluates their effectiveness in the typical book recommendation scenario. We hope that this study can inspire researchers to analyse more opportunities for applying LLMs in similar tasks and further improve their performance in existing scenarios.

\bibliography{main}



 
\end{document}